\documentclass[pre,twocolumn,amsmath,amssymb,floatfix,superscriptaddress,nopacs]{revtex4}

\usepackage{graphicx}
\usepackage{latexsym}
\usepackage{amsmath}
\usepackage{amssymb}
\usepackage{amsfonts}
\usepackage{bm}
\usepackage{color}

\newcommand{\ddiff}{\ensuremath{\mathrm{d}}}
\newcommand{\tincr}{\delta t}
\newcommand{\tsamp}{\Delta t}
\newcommand{\taustar}{t_{\star}}
\newcommand{\Nc}{\ensuremath{N_\mathrm{c}}}

\newcommand{\la}{\left<}
\newcommand{\ra}{\right>}

\newcommand{\TCM}{T_\mathrm{CM}}
\newcommand{\Tglass}{T_{\mathrm{g}}}
\newcommand{\tauhat}{\hat{\tau}}
\newcommand{\taubs}{\tau_{\mathrm{q}}}
\newcommand{\mua}{\mu_{\mathrm{a}}}
\newcommand{\muq}{\mu_{\mathrm{q}}}
\newcommand{\muone}{\mu_1}
\newcommand{\mutwo}{\mu_2}
\newcommand{\muA}{\mu_{\mathrm{A}}}
\newcommand{\muAhat}{\hat{\mu}_{\mathrm{A}}}
\newcommand{\muF}{\mu_{\mathrm{F}}}

\begin{document}

\title{On inequalities of shear modulus contributions in disordered elastic bodies}

\author{J.P.~Wittmer}
\email{joachim.wittmer@ics-cnrs.unistra.fr}
\affiliation{Universit\'e de Strasbourg \& CNRS,
Institut Charles Sadron, 23 rue du Loess, 67034 Strasbourg, France}
\author{H. Xu}
\affiliation{Universit\'e de Lorraine, LCP-A2MC, 1 Bd Arago, 57070 Metz, France}

\begin{abstract}
We investigate generic inequalities of various contributions to the shear modulus $\mu$ 
in ensembles of amorphous elastic bodies. We focus first on a simple elastic network model with 
connectivity matrices (CMs) which are either annealed or quenched, at or out of equilibrium. 
The stress-fluctuation formalism relation for $\mu$ is rewritten as $\mu = \muone + \mua$ 
with $\muone \ge 0$ characterizing the variance of the quenched shear stresses
and $\mua$ being a simple average over all states and CMs. For equilibrium CM-distributions 
$\mua$ becomes equivalent to the shear modulus of annealed systems, i.e. $\mua \ge 0$,
while more generally $\mua$ may become strongly negative
as shown by considering a temperature quench and a scalar active two-temperature model.
Consistent relations are also found for glass-forming colloids
where $\mu-\muone=\mua=0$ for equilibrium ensembles,
i.e.  $\mu$ is set by the quenched shear stresses, 
while $\mua$ becomes again negative otherwise.
\end{abstract}
\date{\today}
\maketitle


\section{Introduction}
\label{sec_intro}

\begin{figure}[t]
\centerline{
\resizebox{.9\columnwidth}{!}{\includegraphics*{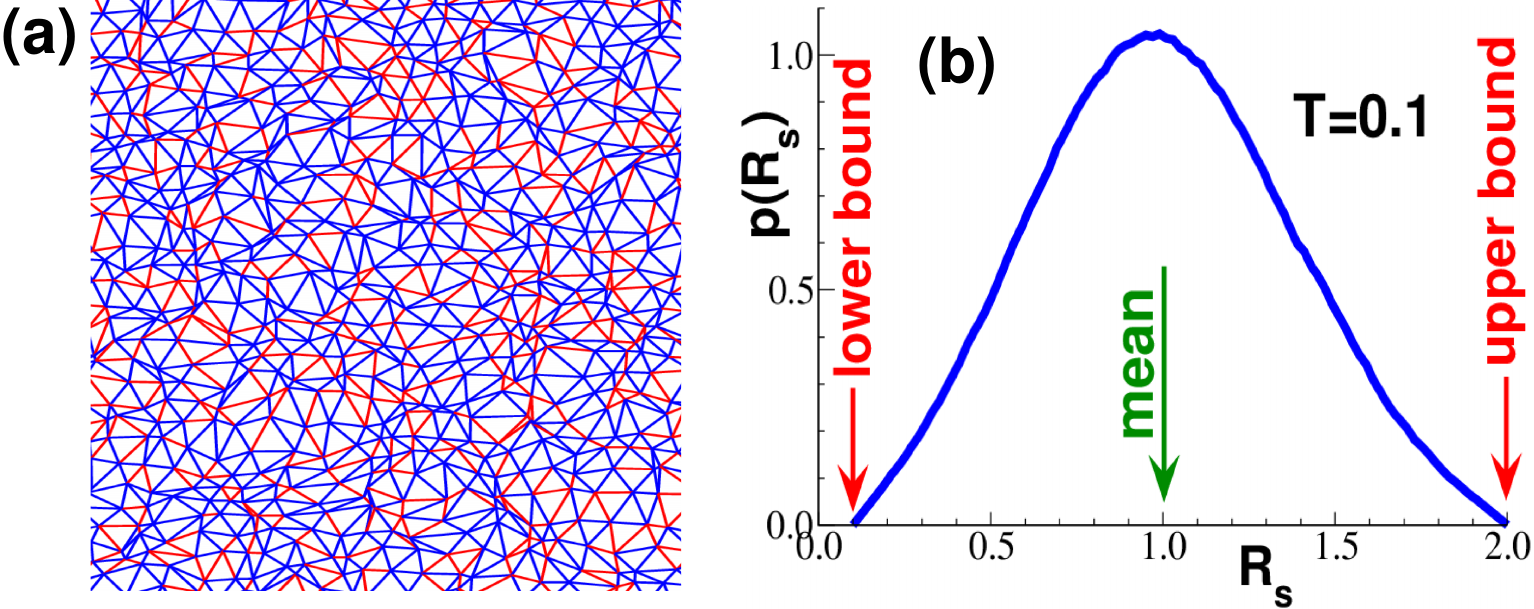}}
}
\caption{Elastic ideal spring network with annealed reference lengths $R_s$ of
the springs imposing $0.1 \le R_s \le 2$ and $\la R_s \ra_s =1$ for a temperature $T=0.1$:
{\bf (a)}
Snapshot of a small subvolume comprising about 500 vertices.
Red lines indicate compressed springs ($r < R_s$), blue lines extended springs.
{\bf (b)}
Normalized distribution $p(R_s)$.
}
\label{fig_model}
\end{figure}

\paragraph*{\bf General context and goal.}
Broken symmetries \cite{Anderson64,ChaikinBook,foot_symbreak} 
may become {\em per\-ma\-nent\-ly} frozen in time (``$\taustar=\infty$")
but they may also be only {\em transient} being annealed 
after some finite characteristic relaxation time $\taustar$.
Permanent or transiently broken
symmetries do not only play a role for crystalline solids \cite{Anderson64,AshcroftMermin} 
but are also crucial for complex matter systems without 
``broken {\em continuous} symmetries" but frozen local heterogeneities 
\cite{DhontGlasses,Barrat06,Pablo03,Pablo04,Barrat09,Barrat13,Barrat14,DelGardoE18,DelGardoE19,spmP5a,ANS24}.
As sketched in Fig.~\ref{fig_model}, 
we investigate in the present study amorphous solids formed by ideal spring networks with 
randomly frozen ``connectivity matrices" (CMs) describing the interactions between the 
connected particles (``vertices") \cite{foot_CM} 
and, more briefly, standard colloidal glasses created by a temperature 
quench from equilibrated high-temperature configurations 
\cite{DhontGlasses,WXP13,Berthier17}. 
Due to the quenched disorder one naturally needs to be more precise concerning the averaging 
procedure of fluctuations \cite{spmP1} being crucial for the determination of second derivatives 
of the free energy \cite{ChaikinBook}. Specifically, we shall focus on the shear modulus $\mu$ 
of amorphous solids computed by means of the stress-fluctuation formalism 
\cite{RowlinsonBook,Hoover69,Lutsko89,Barrat88,WTBL02,Lemaitre04,Pablo05,SBM11,WXP13,WKC16,Procaccia16,lyuda21b,spmP1}.
We thus analyze various ``moments" of the instantaneous shear stress $\tauhat(c,k)$ 
with $k$ denoting the thermodynamic state of a CM $c$. For instance, we denote by 
$\taubs(c) \equiv \langle \tauhat(c,k) \rangle_k$ the $k$-averaged quenched shear stress 
for a given $c$. 
Consistently with the more specific cases discussed elsewhere
\cite{CWBC98b,Liu02,Liu03,ZT13,WKC16,Zaccone19,DelGardoE18,DelGardoE19,lyuda21b},
the rescaled variance $\muone$ of these 
quenched shear stresses is quite generally shown to yield an important contribution 
to the shear modulus $\mu = \muone+\ldots$
This is especially relevant for quenched liquid-type glass-formers under 
equilibrium conditions (as properly defined in Sec.~\ref{sec_liquid}) for which 
we shall demonstrate that
\begin{equation}
0 = \mu - \muone \mbox{ with } \muone = 
\beta V \left[ \langle \taubs(c)^2 \rangle_c - \langle \taubs(c) \rangle_c^2\right]
\label{eq_key_liquid}
\end{equation}
corroborating a recently made suggestion \cite{WKC16}.
($\beta=1/T$ stands for the inverse of the temperature $T$,
$V$ for the $d$-dimensional volume.)

\paragraph*{\bf Outline.}
We begin in Sec.~\ref{sec_theo} by reminding the stress-fluctuation relations 
for the shear modulus and state then general inequalities assuming permanent CMs ($\taustar=\infty$) 
taken from an equilibrium distribution.
The simple elastic network model shown in Fig.~\ref{fig_model} is described in Sec.~\ref{sec_model}.
The static relations of Sec.~\ref{sec_theo} are generalized into the time domain \cite{foot_notation}
by allowing the reorganization of the CMs of the networks, 
i.e. $\taustar(\omega)$ becomes finite with the imposed 
network reorganization frequency $\omega$ being a convenient tuning parameter. 
Computational results obtained for this model are presented in Sec.~\ref{sec_eCMs}
for CMs at equilibrium and in Sec.~\ref{sec_nCMs} for out-of-equilibrium CMs.
We consider especially an initially out-of-equilibrium CM-distribution caused by a 
temperature jump and ``scalar active" networks \cite{Grosberg15,Xu20} 
with different temperatures for the network vertices and the annealed CMs.
In Sec.~\ref{sec_liquid} we turn finally to quenched colloidal glasses and 
confirm that Eq.~(\ref{eq_key_liquid}) must in general hold for quenched ensembles kept at equilibrium. 
We conclude in Sec.~\ref{sec_conc} with an outlook on a similar
relation for the microscopic shear modulus being relevant, e.g., 
for reversible physical gels \cite{WKC16,DelGardoE18,DelGardoE19} or
polymer melts \cite{RubinsteinBook}.

\section{Theoretical considerations}
\label{sec_theo}

\paragraph*{\bf Stress-fluctuation formalism.}
%
Having first been used by Rowlinson for the compression modulus of liquids \cite{RowlinsonBook} and 
later generalized by Squire {\em et al.} \cite{Hoover69} for solid bodies,
the stress-fluctuation formalism provides a convenient route to calculating the 
(isothermic) elastic constants in computer simulations \cite{TadmorMMBook}. 
It has thus been widely used in the past both for macroscopic elastic moduli
\cite{Hoover69,Lutsko89,Barrat88,WTBL02,Lemaitre04,Pablo05,SBM11,WXP13,WKC16,Procaccia16,lyuda21b,spmP1}
as for local mechanical properties
\cite{Lutsko89,Barrat06,Pablo03,Pablo04,Barrat09,Barrat13,Barrat14,spmP5a,ANS24}.
We note that the relations given in the literature depend somewhat on whether the elastic moduli are 
{\em defined} with respect to an unstressed reference state \cite{Hoover69,TadmorMMBook}
or with respect to an in general prestressed reference \cite{RowlinsonBook}.
The difference between both definitions is described by so-called ``Birch coefficients"
as discussed elsewhere \cite{Birch37,Wallace70,SBM11}. 
We take here the latter more physical definition where for gases and liquids
the shear modulus strictly vanishes and the compression modulus is consistent 
with Rowlinson's formula \cite{RowlinsonBook,AllenTildesleyBook,WXP13}.
 
\paragraph*{\bf Shear modulus of amorphous solids.}
We summarize here the salient relations for the macroscopic shear modulus $\mu$
relevant for ensembles of quenched amorphous bodies.
Let $E(c,k)$ denote the energy for a state $k$ of a CM $c$ at a given volume $V$ and shear strain $\gamma$. 
Following Refs.~\cite{WKC16,Procaccia16,spmP1} we expand $E(c,k)$ assuming a canonical 
transformation \cite{spmP1} of the positions and momenta of the particles of the
system under an infinitessimal affine simple shear strain $\gamma$ in, say, the $xy$-plane of the system.
The instantaneous shear stress $\tauhat(c,k)$ and the instantaneous 
affine shear modulus $\muAhat(c,k)$ are then defined by 
\begin{equation}
\tauhat(c,k) \equiv 
\frac{\delta E(c,k)/V}{\delta \gamma} 
\mbox{ and }
\muAhat(c,k) \equiv 
\frac{\delta^2 E(c,k)/V}{\delta \gamma^2}
\label{eq_muAhat_def}
\end{equation}
as the first and the second functional derivatives with respect to this canonical 
transformation taken at $\gamma=0$ for constant $V$ \cite{WKC16}.
While the frozen shear stresses $\taubs(c) \equiv \la \tauhat(c,k) \ra_k$ 
are in general finite, the $c$-ensemble has {\em by symmetry} 
an average shear stress
\begin{equation}
\la \la \tauhat(c,k) \ra_k \ra_c = \la \taubs(c) \ra_c = 0.
\label{eq_avtauus}
\end{equation}
We analyze below the behavior of the (rescaled) moments
\begin{eqnarray}
\muone & \equiv & \la \muone(c) \ra_c 
\mbox{ with } \muone(c) \equiv \beta V \la \tauhat(c,k) \ra_k^2, \label{eq_muone_def}\\
\mutwo & \equiv & \la \mutwo(c) \ra_c 
\mbox{ with } \mutwo(c) \equiv \beta V \la \tauhat(c,k)^2 \ra_k, \label{eq_mutwo_def}\\
\muF   & \equiv & \la \muF(c) \ra_c 
\mbox{ with } \muF(c) \equiv \mutwo(c) - \muone(c), \label{eq_muF_def}\\
\muA   & \equiv & \la \muA(c) \ra_c 
\mbox{ with } \muA(c) \equiv \la \muAhat(c,k) \ra_k \mbox{ and } \label{eq_muA_def}\\
\mu    & \equiv & \la \mu(c) \ra_c \ \
\mbox{ with } \mu(c) \equiv \muA(c) - \muF(c) \label{eq_mu_def}
\end{eqnarray}
being all $\ge 0$. 
As shown elsewhere \cite{WXP13,Procaccia16}, 
the indicated shear modulus $\mu(c)$ for each $c$ is consistent with the 
thermodynamic definition 
\begin{equation}
\mu(c) = \frac{\partial^2 F(T,V,\gamma,c)/V}{\partial \gamma^2},
\label{eq_F2muc}
\end{equation}
i.e. the second derivative of the free energy $F(T,V,\gamma,c)$
taken at $\gamma=0$ at constant $T$, $V$ and $c$.
$\muF$ characterizes the typical shear-stress fluctuations for each $c$
and  $\muA$ the $k$- and $c$-averaged ``affine shear modulus".
Due to Eq.~(\ref{eq_avtauus}) the (rescaled) variance $\muone$
is a measure of the symmetry breaking associated with the quenched shear stresses.
This is mathematically manifested by the fact that
while the $k$- and $c$-averages commute for the ``simple averages" 
\cite{AllenTildesleyBook} $\mutwo$ and $\muA$, 
this is not the case for the ``fluctuation" $\muone$ since in general
\begin{equation}
\langle \langle \tauhat(c,k) \rangle^2_k \rangle_c \ne 
\langle \langle \tauhat(c,k) \rangle^2_c \rangle_k=
\langle \langle \tauhat(c,k) \rangle_k \rangle_c^2=0.
\label{eq_muone_noncommute}
\end{equation}

\paragraph*{\bf Inequalities.}
Being a thermodynamic susceptibility $\mu(c) \ge 0$ must hold for each CM $c$
and, hence, also $\mu \ge 0$ for its $c$-average. Let us rewrite the $c$-average as
\begin{equation}
\mu = \muone + \mua \mbox{ with } \mua \equiv \muA-\mutwo. \label{eq_mua_def}
\end{equation}
Since $\muF \ge 0$ and $\muone \ge 0$ we always have
\begin{equation}
\muA \ge \mu \ge \mua. \label{eq_ineqA}
\end{equation}
What is the meaning of $\mua$? 
Let us denote by $l$ all combined states lumping together all $(c,k)$.
Reminding also Eq.~(\ref{eq_avtauus}), the ``simple average" \cite{AllenTildesleyBook,spmP1}
\begin{equation}
\mua = \la \muAhat(l) - \beta V \tauhat^2(l) \ra_l
\label{eq_mua_simpleaverage}
\end{equation}
would correspond to the stress-fluctuation formula for the shear modulus 
{\em if} the symmetry breaking between different CMs could be ignored,
i.e., to be a thermodynamically meaningful modulus not only 
the states $k$ of the given CMs $c$ must be at thermal equilibrium but also the CMs. 
To see why $\mua \ge 0$ for equilibrium CM-distributions let us assume 
that the broken symmetry becomes lifted by reversible reorganizations of the
CMs for $t \gg \taustar$ assuming (temporarily) a finite $\taustar$.
Using Eq.~(\ref{eq_avtauus}) implies that $\mua = \mu -\muone = \mu$
is the shear modulus of the {\em annealed} CMs and, hence, $\mua \ge 0$.
Interestingly, being a simple average, the expectation value of $\mua$ 
does not change if evaluated for permanently quenched CMs 
of the {\em same} CM-distribution. However, for quenched CMs with $\muone > 0$ 
it then differs from the shear modulus 
$\muq$ due to the inequality $\muq = \mua + \muone > \mua$.
(Instead of $\mu$ we often use below $\muq$ for quenched CMs with $\taustar = \infty$.)
Interestingly, the above argument implies 
\begin{equation}
\mua \ge 0 \Leftrightarrow  \muA \ge \mutwo \Leftrightarrow
\muq \ge \muone \label{eq_ineqB}
\end{equation}
if the (quenched) CMs are taken from an equilibrium distribution.
This is the most central point we want to highlight in this work.
The special limit where the above inequalities reduce to Eq.~(\ref{eq_key_liquid}) 
will be further investigated in Sec.~\ref{sec_liquid}.
Moreover, while Eq.~(\ref{eq_ineqB}) holds for equilibrium CM-distributions 
it may be violated for more general distributions as shown in Sec.~\ref{sec_nCMs}.
We emphasize finally that in the above argument leading to 
Eq.~(\ref{eq_ineqB}) the specific Hamiltonian was irrelevant and there was no need
to specify $\tauhat$ and $\muAhat$ defined by Eq.~(\ref{eq_muAhat_def}).
See Refs.~\cite{Hoover69,Barrat88,Pablo05,WXP13,WKC16,Procaccia16,spmP1,Zaccone19} for specific cases.


\section{Simple elastic network model}
\label{sec_model}

\paragraph*{\bf Springs and topology.}
We use for illustration purposes a simple two-dimensional model elastic network 
where $n$ vertices of mass $m$ are connected by ideal springs $s$ of energy 
$\frac{K}{2} (r-R_s)^2$ with identical spring constants $K$ but a specific reference 
length $R_s$ for each spring. We assume $m=1$, $K=10$ and $\la R_s \ra_s =1$ 
for the first moment of the $R_s$-distribution for each CM $c$.
The network topology is set up using a regular hexagonal lattice of springs 
where each vertex is connected by identical springs of $R_s=1$ to its six next neighbors. 
At variance to the transient spring networks considered in Ref.~\cite{WKC16},
the springs remain permanently connected to the same vertices, 
i.e. in this respect all CMs are quenched and identical. 
We use below $n=10^4$ vertices contained in periodic rectangular simulation boxes.
The  networks are sampled by means of molecular dynamics (MD) using a Langevin thermostat 
\cite{AllenTildesleyBook}. 
For crystalline networks without disorder $\taubs(c)=0$ by symmetry for all $c$ and $T$
and, hence, $\muone=0$ and $\mu=\mua$.
 
\paragraph*{\bf Annealed and quenched disorder.}
Annealed and quenched disorder enters since the reference lengths may fluctuate under 
the additional constraint that $0.1 \le R_s \le 2$.
This is done by means of a Monte Carlo (MC) procedure where the reference lengths 
of two (randomly chosen) neighboring springs are slightly changed subject to a standard 
Metropolis criterion \cite{AllenTildesleyBook} and that both new lengths are 
within the indicated bounds and their sum remains constant. 
(This may be seen as a simple model for branched wormlike micelles with imposed
number of branching points and total branch length \cite{CC90}.)
The algorithm thus mixes MD moves for the vertices at a given CM with MC steps 
altering the Hamiltonian and thus the CM \cite{foot_CM}.  
Time is measured by the MD procedure where we set $\tincr=10^{-3}$ for the Verlet time 
increment \cite{AllenTildesleyBook}. After each $\tincr$ we attempt for a small 
fraction of pairs of springs an MC hopping move to change their reference lengths. 
The reorganization frequency $\omega$ indicated below is proportional to this small fraction:
$\omega=1$ means that a spring is attempted on average once per time unit.
 
\paragraph*{\bf Configurations and data sampling.}
Results obtained for $\omega=0.01$ and running the systems over $t=10^5$ 
are shown in Fig.~\ref{fig_model}.
As can be seen in panel (a), the network becomes very disordered and appears 
macroscopically homogeneous and isotropic. (This visual impression can be corroborated 
using standard pair correlation functions.) 
Panel (b) shows the corresponding probability distribution for $R_s$.
We thus obtain a large ensemble of $\Nc=500$ independent configurations $c$ 
with an equilibrium CM-distribution.
Assuming ergodicity for each CM $k$-averages are replaced by time-averages 
\cite{AllenTildesleyBook}. For instance, we estimate the $k$-average $\taubs(c)$ by computing the 
time-average $\taubs(c,t)$ of the instantaneous shear stresses $\tauhat$ over a ``sampling time" $t$. 
The above-introduced moments thus become {\em apriori} $t$-dependent
as emphasized by an additional argument $t$ \cite{foot_notation}.
We remind that for stationary systems the dynamical shear modulus $\mu(t)$ 
is a smoothing function \cite{spmP1,foot_Gt} 
of the shear stress relaxation modulus $G(t)$ \cite{RubinsteinBook}.
Data are sampled either for strictly permanent networks ($\omega=0$) or by gradually 
reorganizing the CMs at small finite $\omega$ as in our work on transient self-assembled 
spring networks \cite{WKC16,spmP1}.
The behavior of permanent CMs is then obtained from the small-$\omega$ asymptotic limit.

\section{Elastic networks at equilibrium}
\label{sec_eCMs}

\begin{figure}[t]
\centerline{\resizebox{0.9\columnwidth}{!}{\includegraphics*{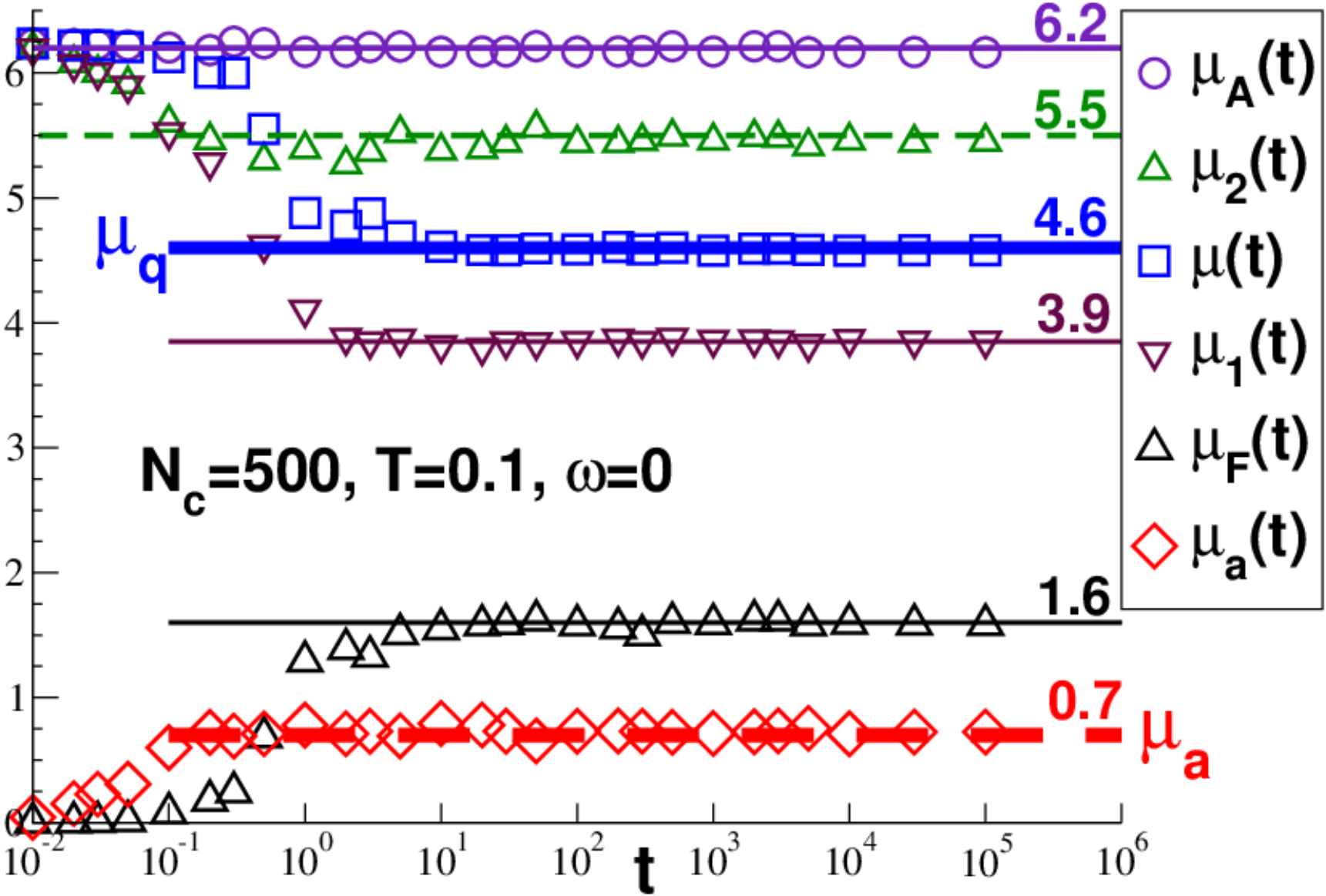}}}
\caption{$\muA(t)$, $\mutwo(t)$, $\mu(t) \equiv \muA(t)-\muF(t)$, $\muone(t)$,  
$\muF(t) \equiv \mutwo(t)-\muone(t)$ and $\mua(t) \equiv \muA(t)-\mutwo(t)$ 
for equilibrated and finally quenched CMs ($\omega=0$). 
Eq.~(\ref{eq_ineqB}) is nicely confirmed.
}
\label{fig_qCM_t}
\end{figure}

\paragraph*{\bf Quenched CM-distributions.}
%
Let us first characterize the moments for switched-off reorganization MC moves 
($\omega=0$, $\taustar=\infty$).
Using a half-logarithmic representation this is shown in Fig.~\ref{fig_qCM_t} 
with the sampling time $t$ being the horizontal axis.
As expected, all moments become rapidly $t$-independent ($t \gg 1$)
as emphasized by horizontal lines. The static asymptotic limits are
\begin{equation}
\muA \approx 6.2 > \muq \approx 4.6 > \muone \approx 3.9 \gg \mua \approx 0.7 > 0.
\label{eq_statvalues}
\end{equation}
The inequalities Eq.~(\ref{eq_ineqA}) and Eq.~(\ref{eq_ineqB}) thus hold as expected. 
Please note that $\mua$ is very small albeit finite. 
\begin{figure}[t]
\centerline{\resizebox{0.9\columnwidth}{!}{\includegraphics*{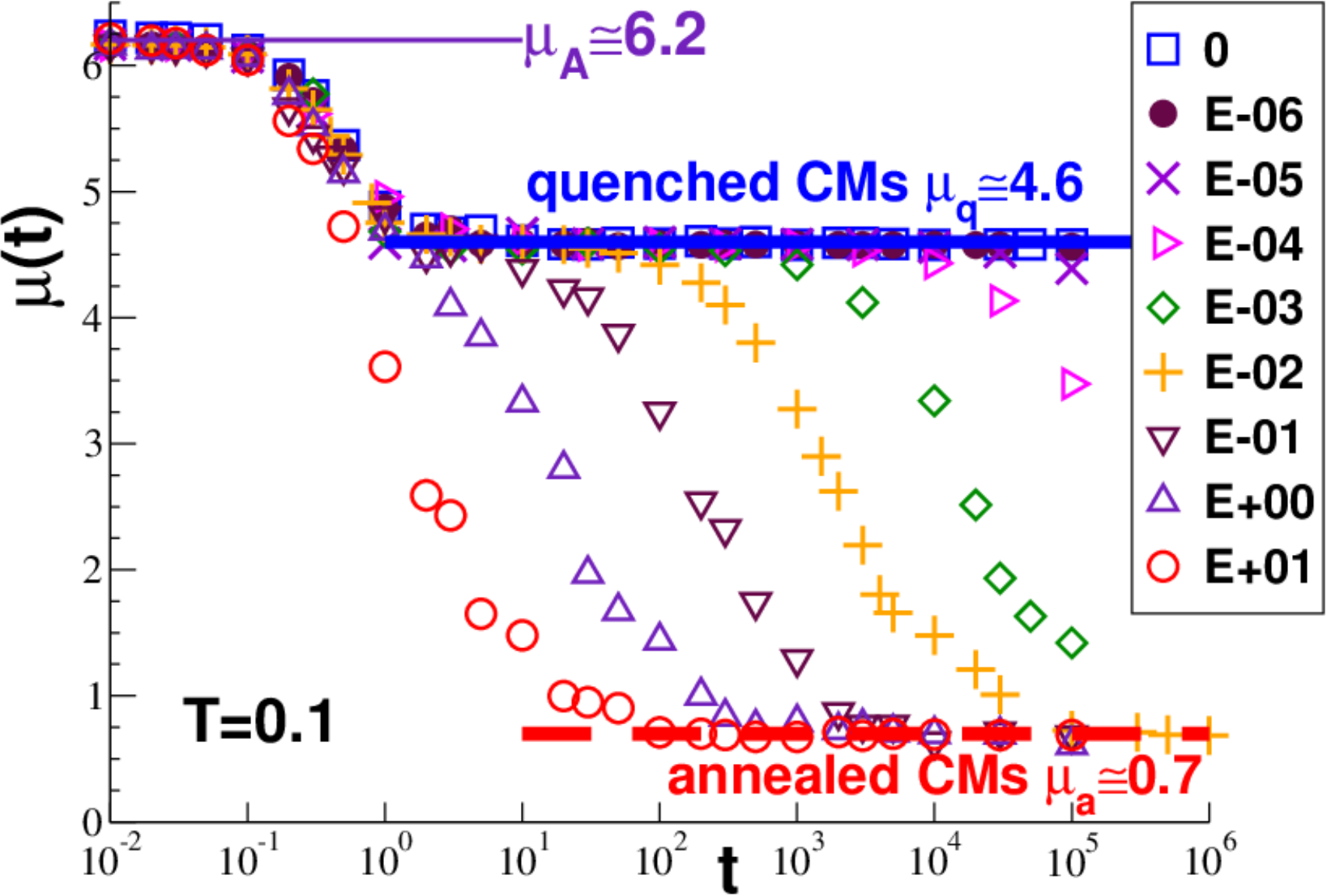}}}
\caption{$\mu(t) \equiv \muone(t) + \mua(t)$ for different $\omega$ as indicated in the legend.
The horizontal lines indicate (top to bottom) the large-$t$ asymptotes 
$\muA \approx 6.2$, $\muq \approx 4.6$ and $\mua \approx 0.7$ for $\omega=0$.
}
\label{fig_mu_t}
\end{figure}

\paragraph*{\bf Finite reorganization rates.}
The thermodynamic static values for permanent CMs are also of relevance 
if the reorganization MC moves are not completely switched off. 
This is demonstrated for a large range of $\omega$ in Fig.~\ref{fig_mu_t} for $\mu(t)$. 
As can be seen, $\mu(t) \to \muA$ for all $\omega > 0$ for very small times (thin horizontal line) 
\cite{foot_notation}. This is expected since $\muF(t)$ must vanish in this limit.
For small $\omega$ and not too large $t$, $\mu(t)$ approaches the static shear modulus $\muq$ 
of quenched networks (bold horizontal line). 
$\muq$ becomes an intermediate upper shoulder (plateau) upon further increasing $\omega$ or $t$. 
Importantly, for sufficiently large $t$ and $\omega$ the dynamical shear modulus $\mu(t)$ 
always approaches $\mua$ (dashed horizontal line), 
i.e. the static shear modulus of completely annealed networks. 
Naturally, this is best seen for our largest reorganization frequencies ($\omega \gg 0.01$). 
We note finally that the relaxation time $\taustar$,
operationally defined by $\mu(t=\taustar)/\mua -1 = 5\%$,
is inversely proportional to $\omega$ 
and that it is possible (not shown) to collapse the data for large $t$ by
tracing $\mu(t)$ as a function of $t/\taustar(\omega)$.

\section{Out-of-equilibrium networks}
\label{sec_nCMs}
\begin{figure}[t]
\centerline{\resizebox{0.9\columnwidth}{!}{\includegraphics*{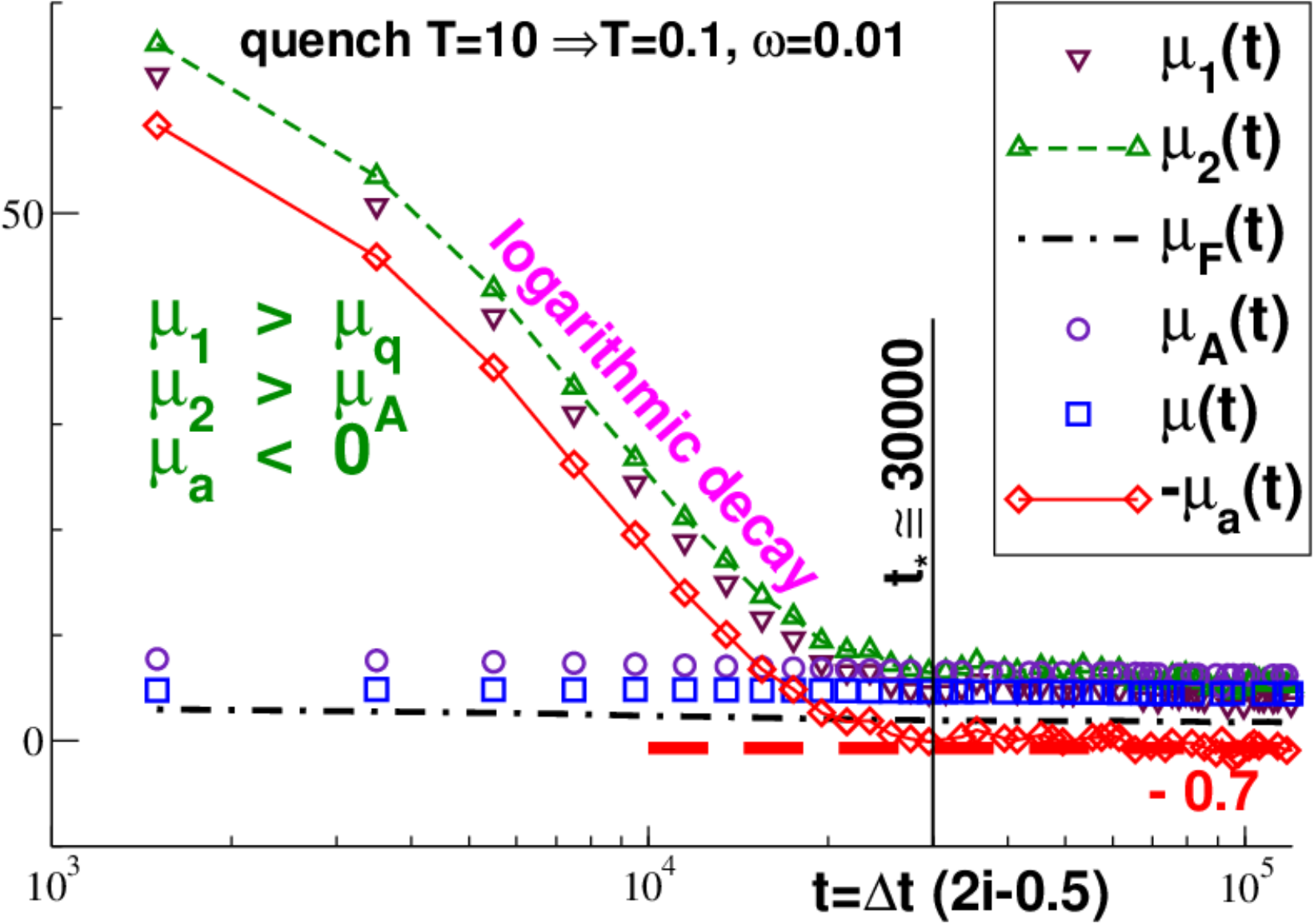}}}
\caption{Time-dependence of various moments after quenching at $t=0$ 
from $T=10$ to $T=0.1$ and allowing the CMs to reorganize with a small frequency $\omega=0.01$. 
Eq.~(\ref{eq_ineqB}) gets violated for $t \ll \taustar \approx 30000$
(vertical line) where $\muone(t)$, $\mutwo(t)$ and $-\mua(t)$ are found to decay logarithmically.
}
\label{fig_Tquench}
\end{figure}

Having confirmed Eq.~(\ref{eq_ineqB}) for equilibrium CMs 
we show now using two examples with non-equilibrium CMs
that these inequalities do not hold in general.

\paragraph*{\bf Temperature jump.}
Starting with annealed liquids at high temperatures amorphous solids are commonly 
created by a rapid quench below the glass transition temperature $\Tglass$ \cite{DhontGlasses}. 
This has not only the effect that the relaxation dynamics is strongly reduced but, 
moreover, that glasses are generally not at thermal equilibrium at $T < \Tglass$, 
i.e. Eq.~(\ref{eq_ineqB}) may not hold. This issue will be addressed in Sec.~\ref{sec_liquid}.
Here we consider first a similar situation where networks are first equilibrated with
finite $\omega$ at $T=10$ and then suddenly at $t=0$ quenched to our standard 
temperature $T=0.1$, i.e. the CM-distribution is initially not at equilibrium. 
As shown in Fig.~\ref{fig_Tquench}, it is then allowed to relax with $\omega=0.01$. 
The data points have been obtained using an iteration of subsequent tempering and measuring 
intervals of both $\tsamp=1000$.
The indicated time for the horizontal axis is $t = \tsamp (2i - 0.5)$ for each iteration step $i$.
While Eq.~(\ref{eq_ineqA}) always holds it is seen that Eq.~(\ref{eq_ineqB}) 
is strongly violated for short times $t \ll \taustar$ with $\taustar \approx 30000$. 
In the large-$t$ limit the equilibrium CM-distribution for $T=0.1$ is restored,
cf.~Eq.~(\ref{eq_statvalues}).

\begin{figure}[t]
\centerline{\resizebox{0.9\columnwidth}{!}{\includegraphics*{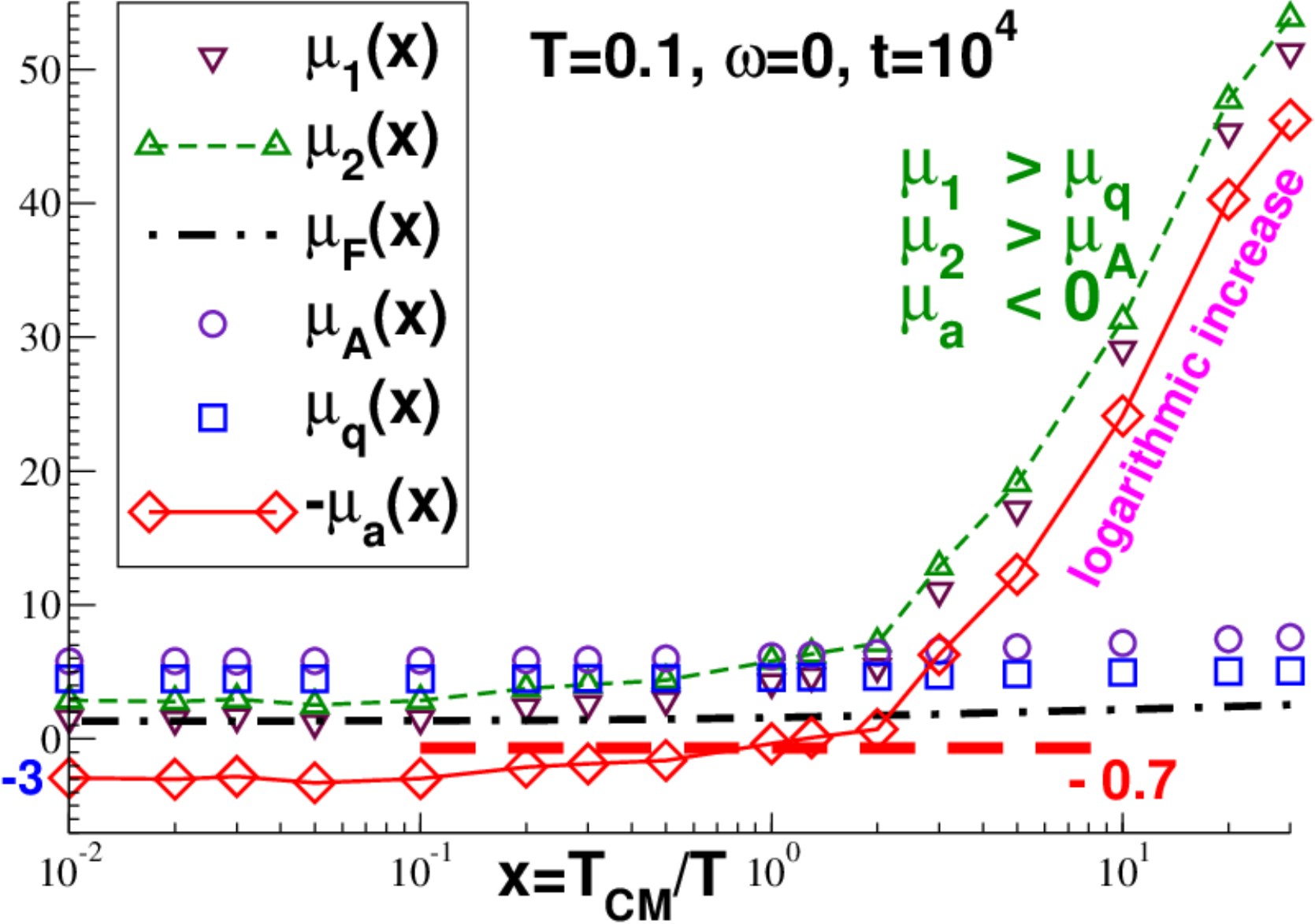}}}
\caption{Scalar active two-temperature model with $T=0.1$ for the network vertices 
and $\TCM$ for the CMs. A half-logarithmic representation of the observables
as a function of the reduced temperature $x=\TCM/T$ is used. 
The final production runs are performed at switched-off MC moves ($\omega=0$) sampling 
over a sampling time $t=10^4$. 
Equation~(\ref{eq_ineqB}) is violated for $x \gg 1$ but holds for $x \ll 1$.
}
\label{fig_Tfalse}
\end{figure}

\paragraph*{\bf Active two-temperature model.}
As a second example we present in Fig.~\ref{fig_Tfalse} data for two-temperature
networks where the MD dynamics of the vertices is coupled to a Langevin thermostat
with $T=0.1$ while a broad range of temperatures $\TCM$ is imposed for the MC steps 
of the network reorganization. This model is motivated by recent work on 
scalar active matter with different temperatures for each particle class
\cite{Grosberg15,Xu20}.  
Half-logarithmic coordinates are used with $x=\TCM/T$ being the dimensionless horizontal axis.
For each $x$ we anneal the CM-ensemble using a finite $\omega$ over sufficiently long times
until all properties become stationary. The reorganization MC steps are then switched off 
for the production runs presented in Fig.~\ref{fig_Tfalse} ($\omega=0$, $\taustar=\infty$, $t=10^4$).
As expected, the data points for $x\approx 1$ are consistent with Fig.~\ref{fig_qCM_t}
and $\muA(x)$, $\muF(x)$ and $\muq(x)$ are seen to only depend very weakly on $x$.
Importantly, we observe for $x \gg 1$ that the inequalities Eq.~(\ref{eq_ineqB}) 
are strikingly violated, just as for the short-time limit in Fig.~\ref{fig_Tquench}.
This makes sense since in both cases the CMs are distributed using a too large temperature
which generates stronger quenched stresses.
Moreover, Eq.~(\ref{eq_ineqB}) does not only hold for $x \approx 1$ but also for $x \ll 1$. 
We also note that in the small-$x$ limit the moments $\muone(x)$ and $\mutwo(x)$,
measuring the quenched shear stresses, slightly decrease albeit remaining finite.
Hence, $-\mua(x)=\mutwo(x)-\muA(x)$ also decreases with $-\mua(x) \to -3$ for $x \to 0^+$.

\section{Liquid-type elastic solids}
\label{sec_liquid}

\paragraph*{\bf General considerations.}
We have considered above elastic networks which by construction must have a {\em finite} shear 
modulus even for annealed reference lengths $R_s$ where $\mu=\mua>0$. 
We turn now to liquid-type systems such as 
reversible physical gels \cite{WKC16,DelGardoE18,DelGardoE19}
or polymer melts \cite{RubinsteinBook} where 
\begin{equation}
\mu \simeq \mua \simeq \muone \simeq 0 \mbox{ for } t \gg \taustar 
\label{eq_liquid_def}
\end{equation}
with $\taustar$ being the finite terminal relaxation time of the system.
Obviously, in many real experiments and computer simulations $\taustar$ becomes rapidly extremely large, 
especially upon cooling, and in practice $t \ll \taustar$ for all realistic sampling times $t$.
One thus expects elastic solid behavior with a finite shear modulus $\mu(t) \approx \muq > 0$
for a finite $t$-window similar to the low-$\omega$ data presented in Fig.~\ref{fig_mu_t}
for elastic networks. Occasionally, such effectively quenched liquid-type systems
may still be at thermal equilibrium \cite{WKC16}. 
(For reversible physical gels this may be achieved, e.g.,
by temporarily adding some catalytic compounds facilitating the network reorganization or 
in computational work by means of additional efficient albeit artificial MC moves 
as shown below for one example.)
Assuming thus an equilibrium distribution of dynamically quenched configurations,
the inequalities Eq.~(\ref{eq_ineqB}) are expected to become {\em equalities}, i.e.
\begin{equation}
0 = \mua \equiv \muA-\mutwo= \muq - \muone.
\label{eq_liquid}
\end{equation}
In agreement with related recent studies \cite{WKC16,lyuda21b,spmP1}
we have thus finally confirmed Eq.~(\ref{eq_key_liquid}).
Note that failure of Eq.~(\ref{eq_liquid}) implies that the quenched 
configuration ensemble is inconsistent with an equilibrium distribution.

\begin{figure}[t]
\centerline{\resizebox{0.9\columnwidth}{!}{\includegraphics*{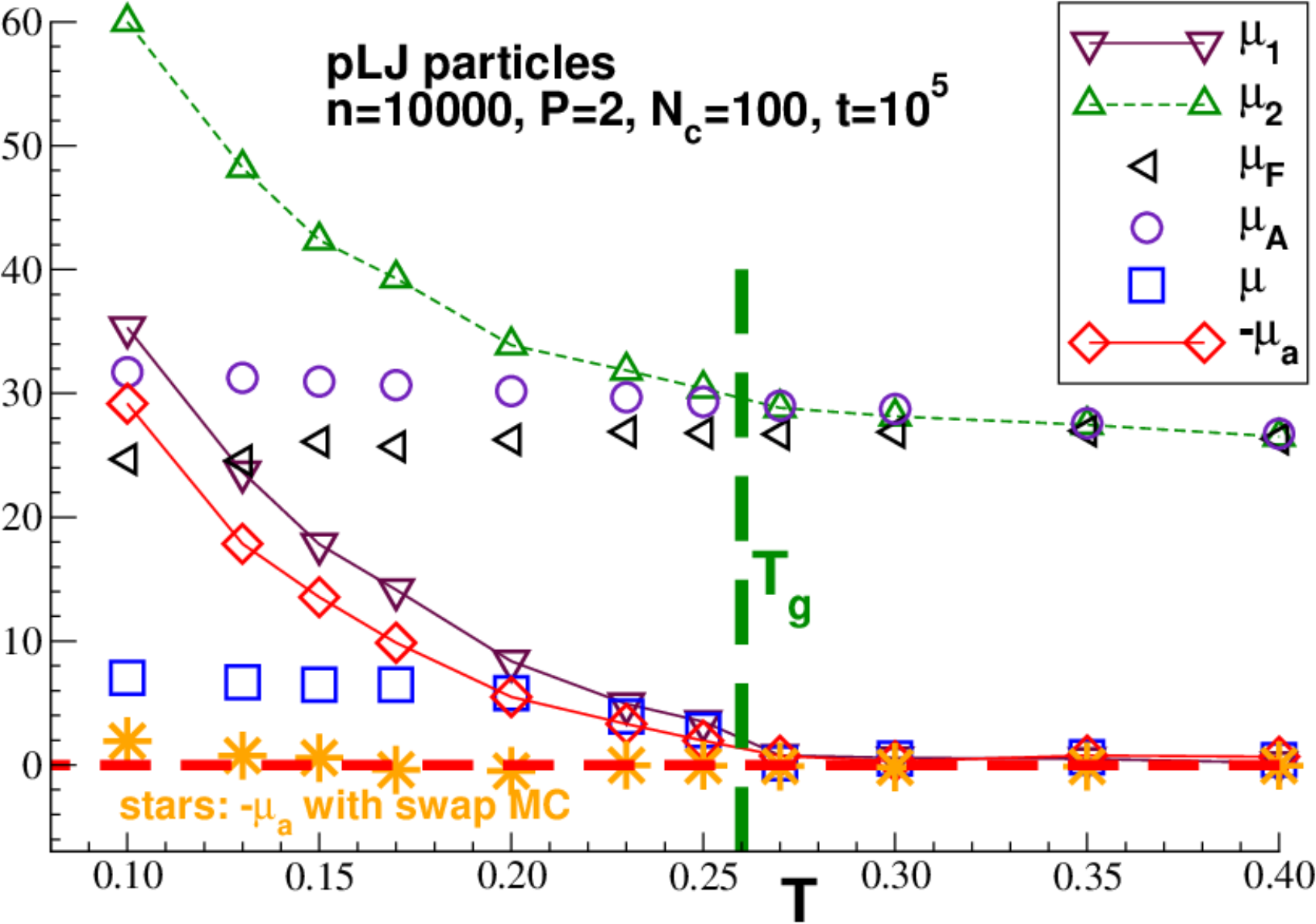}}}
\caption{$T$-dependence of contributions to the shear modulus 
for $n=10000$ pLJ particles at an imposed average normal pressure $P=2$ 
\cite{WXP13,spmP1}. 
All data points are obtained using MD-production runs over $t=10^5$
and averaged over $\Nc=100$ independent configurations. 
Open symbols refer to configurations obtained by means of a constant-rate $T$-quench
while the stars show $-\mua$ for configurations being first thoroughly
equilibrated using a mix of local and swap MC moves \cite{Berthier17,spmP1}.
While Eq.~(\ref{eq_liquid}) is violated in the first case below $\Tglass \approx 0.26$ 
(vertical dashed line) it holds in the second case even for $T \ll \Tglass$
(horizontal dashed line).}
\label{fig_liquid}
\end{figure}

\paragraph*{\bf Colloidal glass-former.}
The above statements are illustrated in Fig.~\ref{fig_liquid} 
using systems of polydisperse Lennard-Jones (pLJ) particles which have been
extensively investigated in the past \cite{WTBL02,WXP13,lyuda21b,spmP1}. 
We average over $\Nc=100$ independent configurations which are kept at an average 
normal pressure $P=2$ by means of a standard barostat \cite{AllenTildesleyBook}.
All production runs are performed using straight-forward MD (with switched off barostat)
just as for the spring networks investigated above. We only present data for $t=10^5$. 
Two different cases of ensemble preparation are compared.
The open symbols in Fig.~\ref{fig_liquid} refer to configurations created starting from 
equilibrated start configurations at $T=0.5$ using a standard quench protocol decreasing
the imposed temperature with a constant quench rate $0.5 \cdot 10^{-5}$ (in LJ units)
and a final tempering over $t=10^5$ for each $T$ sampled.
In addition, we thoroughly equilibrated each configuration  
using a mix of local and swap MC moves over $10^6$ MC cycles 
\cite{Berthier17,spmP1} and sample this second ensemble again using MD.
For clarity, we only show for this second case data for $-\mua$ (stars).
This shows that for well-equilibrated systems Eq.~(\ref{eq_liquid})
holds in principle for all temperatures.
(As revealed by closer inspection, $\mua = 0$ is slightly violated for our lowest 
temperatures suggesting that additional equilibration with swap MC moves is warranted.)
This is different for the first data set (open symbols) obtained using the standard MD-quench 
for which Eq.~(\ref{eq_liquid}) is clearly violated below the glass-transition
temperature $\Tglass \approx 0.26$ \cite{WXP13,lyuda21b}.
Especially, it is seen that $\muone \gg \mu$ and $-\mua=\mutwo-\muA \gg 0$. 
We thus found similar behavior as for the networks in Fig.~\ref{fig_Tquench}.

\section{Conclusion}
\label{sec_conc}

\paragraph*{\bf Summary.}
We have investigated in this work general inequality relations for the 
different contributions to the shear modulus of amorphous bodies with annealed or 
quenched configurations.
We compared ensembles of quenched configurations at thermal equilibrium with more 
general distributions with a different dispersion of quenched shear stresses.
To illustrate our key points we have first focused on simple elastic spring
networks for which the disordered CMs can be simply defined and operationally
changed in terms of the reference lengths $R_s$ of the ideal springs, cf.~Fig.~\ref{fig_model}.
The stress-fluctuation formalism relation for the shear modulus was rewritten as
$\mu = \muone + \mua$ with $\muone \ge 0$ being the (rescaled) variance of the quenched shear stresses
and $\mua$ a simple average over all states and CMs, cf.~Eq.~(\ref{eq_mua_simpleaverage}).
For annealed networks $\mua$ is equivalent to the shear modulus $\mu$ as shown in Fig.~\ref{fig_mu_t}. 
Interestingly, $\mua$ may be strongly negative for out-of-equilibrium distributions 
as shown by considering a temperature quench (cf.~Fig.~\ref{fig_Tquench}) and
a scalar active two-temperature model (cf.~Fig.~\ref{fig_Tfalse}).
While the shear modulus can still be computed in these cases
using the stress-fluctuation relation Eq.~(\ref{eq_mua_def}),
the more special inequalities Eq.~(\ref{eq_ineqB}) do not hold in general
and $\muone$ may thus differ strongly from the shear modulus.
Turning in Sec.~\ref{sec_liquid} to glass-forming pLJ particles
we have confirmed that similar behavior can be expected for other types of
amorphous elastic solids. While Eq.~(\ref{eq_liquid}) was seen to hold for
pLJ glasses equilibrated by means of the swap MC algorithm, systematic 
and strong deviations arise for standard out-of-equilibrium ensembles prepared
using a conventional quench protocol (cf.~Fig.~\ref{fig_liquid}).
The measurement of deviations from Eq.~(\ref{eq_liquid}) for liquid-type systems provides thus
quite generally a convenient diagnosis tool of the equilibration status of configuration ensembles.

\paragraph*{\bf Outlook.}
Further work will focus on the systematic analysis of effects 
related to the finite sampling time $t$ used for our pLJ systems and the characterization 
of the equilibration of the configuration ensemble by tempering with different frequencies 
of swap MC moves. Relaxation behavior similar to Fig.~\ref{fig_Tquench} is expected.
It may also be of interest to consider scalar active self-assembled networks
of colloids bridged by ideal springs \cite{WKC16}
and polydisperse particles with different temperatures as in Refs.~\cite{Grosberg15,Xu20}.
We expect Eq.~(\ref{eq_liquid}) to be violated just as for the two-temperature active network 
presented in Fig.~\ref{fig_Tfalse}. 
The above work has focused on macroscopic observables.  
We note finally that similar relations should hold for the {\em microscopic} shear modulus 
and its contributions at finite wavevectors $q$ \cite{spmP5a,ANS24}.
For melts of long reptating polymer chains at equilibrium it is, e.g., expected
that Eq.~(\ref{eq_key_liquid}) remains valid in reciprocal space 
for sampling times below the reptation time \cite{RubinsteinBook}.
It then follows from general considerations of isotropic tensor fields \cite{spmP5a} 
that the spatial correlations of the (transiently) quenched shear stresses must be long ranged 
decaying as $1/r^3$ with the distance $r$ and the prefactor being set by the intermediate 
plateau $G(t) \approx \mu$ of the shear stress relaxation function.
Work in this direction is currently underway.


\vspace*{0.1cm}
\paragraph*{Acknowledgments.}
We are indebted to J.~Baschnagel (Strasbourg) for helpful discussions.

 
\vspace*{0.1cm}
\paragraph*{Data availability statement.}
Data sets are available from the corresponding author on reasonable request.



\begin{thebibliography}{45}
\bibitem[{foo({\natexlab{a}})}]{foot_symbreak}
\bibinfo{note}{Equations with a particular symmetry can have solutions
  violating this symmetry, although the set of all solutions is symmetrical.
  This is the mathematical essence of ``spontaneous symmetry breaking" in
  theoretical physics including condensed matter physics
  \cite{Anderson64,ChaikinBook}.}
\bibitem[{foo({\natexlab{b}})}]{foot_CM}
\bibinfo{note}{The CM contains information on both the network topology
  specifying which vertex pairs are interacting and the specific interaction
  (Hamiltonian) between two vertices. Only the latter information may be
  annealed in this work.}
\bibitem[{foo({\natexlab{c}})}]{foot_notation}
\bibinfo{note}{We write $f(t)$ if $t$ matters but drop the argument if we focus
  on the large-$t$ limit or an intermediate plateau.}
\bibitem[{foo({\natexlab{d}})}]{foot_Gt}
\bibinfo{note}{It can be shown that $G(t) = \frac{\ddiff^2}{\ddiff t^2} [t^2
  \mu(t)/2]$ for stationary stochastic processes \cite{spmP1}.}
\end{thebibliography}

\end{document}